\documentclass[aps,prb,twocolumn,showpacs,superscriptaddress]{revtex4}
\usepackage{amssymb}
\usepackage{graphicx}

\begin{document}

\title{Bragg-Cherenkov resonance and polaron-like decoupling of the Wigner solid on superfluid helium}
\author{Yu.P. Monarkha}
\email[E-mail: ]{monarkha@ilt.kharkov.ua}
\affiliation{B. Verkin Institute for Low Temperature Physics and Engineering of the National Academy of Sciences of Ukraine, 47 Nauky Ave.,
Kharkiv 61103, Ukraine}

\begin{abstract}
Nonlinear polaron-like dynamics of the two-dimensional Wigner solid (WS) on superfluid $^{4}\mathrm{He}$
are theoretically analyzed in different models and transport regimes for their similarities and distinctions.
The Bragg-Cherenkov (BC) resonant excitation of surface waves and WS decoupling
from surface dimples were usually considered in terms of a dc transport model. At the same time,
field-velocity characteristics of the WS are measured under ac conditions and
presented for time-averaged quantities. Here the nonlinear equation of motion of the WS
coupled to surface dimples is studied for ac conditions using two different approaches
based on fixing the driving field or the output current. Both approaches are shown to give similar
results for the first harmonics of major transport properties. In the ac theory, the BC-resonances
for dimple inertia and the momentum relaxation rate have asymmetrical shapes, which
is in contrast with the results of dc models.
Even a quite low driving frequency is shown to affect the amplitude of the
BC resonance and decoupling of the WS. Above the BC threshold, the effective mass of surface
dimples as a function of the velocity amplitude strongly oscillates indicating
multiple recoupling processes.

\end{abstract}




\maketitle

\section{Introduction}

Transport properties of a two-dimensional (2D) Wigner solid
on the free surface of superfluid helium have much in common with the
polaron dynamics~\cite{MonKon-book}. Though electron interaction with
capillary wave excitations (ripplons) usually is rather weak and the
single-electron self-trapping can happen only at ultra-low temperatures,
in the Wigner solid (WS) state surface dimples appearing under each electron
play a crucial role in the WS dynamics at usual liquid-helium temperatures.
Such a dimple lattice (DL) does
not affect the WS phase transition because the electron-dimple interaction
energy~\cite{MonShi-1975} is much smaller than the typical Coulomb interaction energy
$e^{2}\sqrt{n_{e}}$ (here $n_{e}$ is the areal density of electrons).
Still, the DL affects electron transport along the surface because the effective
mass of a dimple $M_{D}$ is much larger than the free electron mass
$m_{e}$. The linear dynamics of the coupled WS-DL
system~\cite{FisHalPla-1979} had given the possibility to
detect~\cite{GriAda-1979} the WS of surface electrons and to establish
its structure.

Experimental study of the WS conductivity over the liquid helium surface had
revealed a body of remarkable nonlinear phenomena. Firstly, we would like to
note a sharp change in the magnetoresistance of surface electrons~\cite{GiaWil-1991}
attributed to the nonequilibrium melting of the electron
crystal. A similar effect presumably of different origin was reported~\cite{ShiKon-1995}
as an abrupt jump of the ac Corbino conductivity $\sigma _{xx}$.
A mobility jump was observed also in a zero magnetic field~\cite{SyvDotKov-1998}.

By now several theoretical models were proposed to explain the nonlinear conductivity of
the WS on superfluid liquid helium. The magneto-conductivity
jump observed was attributed to the collective sliding of the electrons out
of the periodic deformation of the helium surface (the sliding
model)~\cite{ShiKon-1995,ShiKon-1996}. The Bragg-Cherenkov (BC) scattering
model~\cite{DykRub-1997} describes the saturation of velocity
data~\cite{KriDjeFoz-1996} observed when the Hall velocity $\mathrm{v}_{\mathrm{H}}$
approaches the phase velocity of the capillary wave whose wavevector $\mathbf{q}$
coincides with the smallest reciprocal lattice vector $\mathbf{g}$. Another
dc model explaining the Hall-velocity limited magnetoconductivity~\cite{KriDjeFoz-1996}
and the conductivity jumps~\cite{ShiKon-1995} is based on the
hydrodynamic approach~\cite{Vin-1999}. In this treatment, the BC effect
leads to a resonance increase of the DL depth, and instability
appears as a breach of the balance of forces acting on the WS: at a
certain velocity, electrons decouple from the DL (the decoupling
model). Even though both sliding and decoupling models sometimes are
assumed to be physically the same, we would like to emphasize their
important distinctions. In the sliding model, an electron overcomes the
potential (induced by the dimple) as a function of motion coordinate.
In the decoupling model~\cite{Vin-1999}, the potential induced by the
dimple is considered as a function of electron velocity, while the
relative displacement of the WS and DL is set to zero.

The above noted theoretical models are proposed for dc transport conditions,
while the experiments~\cite{ShiKon-1995,KriDjeFoz-1996,SyvDotKov-1998} are
conducted under ac conditions when the driving electric field $E$ and
electron velocity $\mathit{v}$ are periodic functions of time $t$. The real
quantities measured in this case are somehow averaged or represent
amplitudes of the field and current. A finite frequency should
affect the BC response of liquid media because an electron is
in the BC resonance condition for only a limited time. Under ac conditions with a given
driving electric field $E\left( t\right) =E_{a}\sin \omega t$ (we shall call
it the given-field regime), the equation of motion describing the coupled
WS-DL system is a very complicated nonlinear integro-differential equation.
The accurate solution of this equation was obtained~\cite{MonKon-2009,Mon-2017,Mon-2018}
only for the inverse problem, finding $E\left( t\right) $ for a given
current $j=j_{a}\sin \omega t$ (the given-current regime). This approach
surprisingly leads to the bistability of velocity-field characteristics for
amplitudes $j_{a}$ and $E_{a}$ which can explain mobility jumps observed. It
should be noted here that the dc-decoupling model~\cite{Vin-1999} can be
considered as a particular case of the given-current regime with $j=%
\mathrm{const}$. Under magnetic field directed perpendicular to the surface,
calculated $\sigma _{xx}$ also exhibits a strong jump and a sharp fall which
leads to negative conductivity effects~\cite{Mon-2018}. Qualitatively, the
efficiency of the inverse solution can be
attributed to the fact that in experiments with the WS on liquid helium the
driving electric field is often adjusted to the current owing to electron
redistribution which screens external potential variations. Nevertheless, by
now it was not clear if the effects obtained by solving the inverse problem
can correspond to the solution of the nonlinear equation of motion
for a given driving field $E\left( t\right) $. This question recently
becomes of special interest because of new time-resolved
measurements on WS transport over the surface of superfluid helium confined in a
micron-scale channel~\cite{ReeBeyKon-2016,ReeSheKon-2020,ZouKonRee-2021}.
Additional interest in this problem was inspired by similarities between the
nonlinear transport properties in the WS-DL system and polaron
systems~\cite{JohSta-2004,GeWonLin-1998,GaaKueRei-2007}.

In this work, we theoretically analyze different models and approaches used
for the description of the nonlinear transport of the WS over the free surface
of superfluid helium and compare the results. In particular, we shall show
that the accurate treatment of the given-field regime, $E\left( t\right)
=E_{a}\sin \omega t$, in the lowest nonlinear approximation confirms the
validity of the exact solutions for the dimple-mass $M_{D}$ and the
electron-dimple momentum relaxation rate $\nu _{D}$ found using the given-current regime~\cite{Mon-2018}.
Therefore, the relationship between amplitudes of the first harmonics of the
driving electric field and current obtained in the given-current treatment
can be used for explaining experimental results regardless of the actual
transport regime. Moreover, concentrating on the low-frequency limit ($%
\omega \ll \omega _{g}$, here $\omega _{q}$ is the ripplon spectrum) we
found that the effective mass of surface dimples increases with the velocity amplitude at a
rate which differs from the rate obtained in the DC models. At the same
time, the position and height of the BC-resonance maxima for the dimple mass and momentum relaxation
rate are shown to be strongly affected even by a very low frequency usually used in experiments
on WS transport.

\section{WS-DL coupling: Nonlinear dynamics}

Consider a 2D electron lattice on the free surface of superfluid $^{4}%
\mathrm{He}$ at low temperatures $T<0.4\,\mathrm{K}$. In this case, electrons
interact mostly with surface excitations (ripplons) which have
capillary-wave dispersion $\omega _{q}=\sqrt{\alpha /\rho }q^{3/2}$, here $%
\alpha $ and $\rho $ are the surface tension and mass density of liquid
helium, respectively. The DL is described by a periodic surface distortion $%
\xi \left( \mathbf{r}\right) $. We shall consider only the spatially uniform
driving electric field $E\left( t\right) $ and the WS displacement $u\left(
t\right) $ directed along the $x$-axis. Then, in the coupled equations of motion for the
Fourier transforms $\xi _{\mathbf{q}}$ (only $\mathbf{q}=\mathbf{g}$ are important)
and $u\left(t\right) $, the interaction force can be averaged over fast thermal
and zero-point vibrations~\cite{MonSyv-2012}. Finally, equations describing the WS-DL coupling can
be presented in the following way
\begin{equation}
\ddot{\xi}_{\mathbf{g}}+2\gamma _{g}\dot{\xi}_{\mathbf{g}}+\omega
_{g}^{2}\xi _{\mathbf{g}}=-\frac{n_{e}g}{\rho }\tilde{V}_{g}\mathrm{e}%
^{-ig_{x} u\left( t\right) }  \label{e1}
\end{equation}%
\begin{equation}
m_{e}\ddot{u}\mathbf{+}\sum_{\mathbf{g}}\xi _{\mathbf{g}}\tilde{V}%
_{g}ig_{x}e^{ig_{x} u\left( t\right)}=eE  \label{e2}
\end{equation}%
where $\tilde{V}_{q}=V_{q}\exp \left( -q^{2}\left\langle u_{\mathrm{f}%
}^{2}\right\rangle /4\right) $, $V_{q}$ is the electron-ripplon coupling
function defined in Ref.~\onlinecite{MonKon-book} ($V_{q}\rightarrow eE_{\bot }$ in
the limit of strong pressing electric fields $E_{\bot }$), $n_{e}$ is the
areal electron density, $\left\langle u_{\mathrm{f}}^{2}\right\rangle $ is
the mean-square displacement due to fast modes, and $\gamma _{g}$ is the
damping coefficient which is extremely small for liquid $^{4}\mathrm{He}$.
For typical $n_{e}$ and $0.4\,\mathrm{K}>T>0.1\,\mathrm{K}$, the ratio $\gamma _{g}/\omega _{g}$
varies from $10^{-4}$ to $10^{-7}$.
The Debye-Waller factor appears in $\tilde{V}_{q}$ as a result of averaging
over fast vibrations and a self-consistent procedure (for a review see Ref.%
~\onlinecite{MonSyv-2012}).

Eq.~(\ref{e1}) represents a linear oscillator equation affected by an
electron pressure depending on $u\left( t\right) $. It can be solved
quite generally, and the solution $\xi _{\mathbf{g}}\left( t\right) $ can be
represented in the integral form~\cite{MonKon-2009}:
\[
\xi _{\mathbf{g}}\left( t\right) =-\frac{n_{e}g\tilde{V}_{g}}{\rho \hat{%
\omega}_{g}}\int\limits_{0}^{\infty }\mathrm{e}^{-\gamma _{g}\tau }\sin %
\left( \hat{\omega}_{g}\tau \right) \mathrm{e}^{-ig_{x}u\left( t-\tau
\right) }d\tau ,
\]
where $\hat{\omega}_{g}^{2}=\omega _{g}^{2}-\gamma _{g}^{2}$.
Then, inserting $\xi _{%
\mathbf{g}}\left( t\right) $ into Eq.~(\ref{e2}) one can
find the final nonlinear equation for $u\left( t\right) $%
\begin{equation}
m_{e}\ddot{u}=F_{D}\left( u\right) +eE\left( t\right) ,  \label{e3}
\end{equation}%
where
\[
F_{D}\left( u\right) =-\sum_{\mathbf{g}}\frac{n_{e}g\tilde{V}_{g}^{2}}{%
m_{e}\rho \hat{\omega}_{g}}g_{x}\times
\]%
\begin{equation}
\times \int\limits_{0}^{\infty }\sin \left( \hat{\omega}_{g}\tau \right)
\mathrm{e}^{-\gamma _{g}\tau }\sin \left\{ g_{x}\left[ u\left( t\right)
-u\left( t-\tau \right) \right] \right\} d\tau .  \label{e4}
\end{equation}%
The functional $F_{D}\left( u\right) $ represents the average force
acting on an electron due to the presence of the DL. If the WS is at rest, this force
is obviously zero. When the WS moves along the surface, $F_{D}\left(
u\right) \neq 0$ because of a finite ripplon damping (the dimple shape
becomes asymmetrical along the $x$-axis) and the dimple inertia. In the
general case, the equation of motion of the coupled WS-DL system represents a
nonlinear integro-differential equation whose solution is very difficult to
find.

\subsection{The dc modeling of nonlinear effects}

It is instructive to consider firstly the dc transport model. In this case, $%
u\left( t\right) =\mathrm{v}t$ with $\mathrm{v}=\mathrm{const,}$ and the
integral of Eq.~(\ref{e4}) can be evaluated analytically. The dimple inertia
can't enter the equation of motion of the dc model ($\ddot{u}=0$),
therefore, it is convenient to rewrite the average force acting on an
electron due to the polaronic effect as
\begin{equation}
F_{D}^{\left( \mathrm{dc}\right) }\left( \mathrm{v}\right) =-m_{e}\nu
_{D}^{\left( \mathrm{dc}\right) }\left( \mathrm{v}\right) \mathrm{v},
\label{e5}
\end{equation}%
where $\nu _{D}^{\left( \mathrm{dc}\right) }\left( \mathrm{v}\right) $ plays
the role of a momentum relaxation rate caused by the interaction with a
surface dimple%
\begin{equation}
\nu _{D}^{\left( \mathrm{dc}\right) }\left( \mathrm{v}\right) =\sum_{\mathbf{%
g}}\Gamma _{g}\frac{g_{x}^{2}}{g^{2}}\omega _{g}\mathcal{N}_{\mathrm{dc}%
}\left( \frac{\mathrm{v}}{\mathrm{v}_{g_{x}}},\frac{\gamma _{g}}{\omega _{g}}%
\right) ,  \label{e6}
\end{equation}%
\begin{equation}
\Gamma _{g}=\frac{n_{e}g^{3}\tilde{V}_{g}^{2}}{m_{e}\rho \omega _{g}^{4}},
\label{e7}
\end{equation}%
\begin{equation}
\mathcal{N}_{\mathrm{dc}}\left( \mathrm{v}^{\prime },\gamma _{g}^{\prime
}\right) =\frac{2\gamma _{g}^{\prime }}{\left[ 1-\left( \mathrm{v}^{\prime
}\right) ^{2}\right] ^{2}+4\left( \gamma _{g}^{\prime }\mathrm{v}^{\prime
}\right) ^{2}},  \label{e8}
\end{equation}%
and we have used the following notations: $\mathrm{v}_{g_{x}}=\omega
_{g}/g_{x}$, $\mathrm{v}^{\prime }=\mathrm{v}/\mathrm{v}_{g_{x}}$, $\gamma
_{g}^{\prime }=\gamma _{g}/\omega _{g}$. The dimensionless parameter $\Gamma
_{g}$ represents the strength of the WS-DL coupling. Under usual conditions ($%
n_{e}\approx 5\cdot 10^{8}\,\mathrm{cm}^{-2}$) it is of the order of $10^{2}$.
Actually the result given in Eqs.~(\ref{e5})-(\ref{e8}) is a 3D extension of
the 1D decoupling model~\cite{Vin-1999}. The equation of motion is reduced to
a simple balance equation for the two forces acting on an electron: $%
F_{D}^{\left( \mathrm{dc}\right) }\left( \mathrm{v}\right) +eE=0$, where the
function $F_{D}^{\left( \mathrm{dc}\right) }\left( \mathrm{v}\right) $ has a
maximum when $\mathrm{v}/\mathrm{v}_{\left\vert g_{x}\right\vert }$ is close
to $1$. If $eE$ exceeds the maximum value of $F_{D}^{\left( \mathrm{dc}%
\right) }\left( \mathrm{v}\right) $, the balance of forces is impossible and
the WS decouples from the DL.

The appearance of the BC resonance of $\mathcal{N}_{\mathrm{dc}}\left(
\mathrm{v}^{\prime },\gamma _{g}^{\prime }\right) $ and $\nu _{D}^{\left(
\mathrm{dc}\right) }\left( \mathrm{v}\right) $ is quite obvious because the
substitution $u\left( t\right) =\mathrm{v}t$ makes the right side of Eq.~(\ref%
{e1}) a harmonic force with a frequency $g_{x}\mathrm{v}$ acting on a linear
oscillator with its own frequency $\omega _{g}$. Therefore, in the dc model,
the BC resonance is similar to the usual resonance appearing when the frequency
of an external force $\omega \rightarrow \omega _{g}$.

It should be noted that the dimple effective mass does not enter the balance
of forces of the dc model. In the real case, to reach the decoupling point
one has to increase the driving force and, therefore, it becomes a function
of time $eE\left( t\right) $. Generally, the electric field can be increased
so fast that the DL will not follow the WS, and one can imagine the WS
sliding caused by a relative displacement of WS and DL. This is why we
prefer to separate the decoupling model from the sliding model.

The effective mass of the DL, $N_{e}M_{D}^{\left( \mathrm{dc}\right) }\left(
\mathrm{v}\right) $, can be found as the associated mass induced by the
hydrodynamic velocity field in liquid helium. Using the solution $\xi _{%
\mathbf{g}}\left( t\right) $ mentioned above, one can find $M_{D}^{\left(
\mathrm{dc}\right) }\left( \mathrm{v}\right) $ as%
\begin{equation}
\frac{M_{D}^{\left( \mathrm{dc}\right) }\left( \mathrm{v}\right) }{m_{e}}%
=\sum_{\mathbf{g}}\Gamma _{g}\frac{g_{x}^{2}}{g^{2}}\mathcal{M}_{\mathrm{dc}%
}\left( \frac{\mathrm{v}}{\mathrm{v}_{g_{x}}},\frac{\gamma _{g}}{\omega _{g}}%
\right) ,  \label{e9}
\end{equation}%
where
\begin{equation}
\mathcal{M}_{\mathrm{dc}}\left( \mathrm{v}^{\prime },\gamma _{g}^{\prime
}\right) =\frac{1}{\left[ 1-\left( \mathrm{v}^{\prime }\right) ^{2}\right]
^{2}+4\left( \gamma _{g}^{\prime }\mathrm{v}^{\prime }\right) ^{2}}
\label{e10}
\end{equation}%
is a dimensionless function. Thus, the dimple mass increases with $\mathrm{v}
$ similarly to $\nu _{D}^{\left( \mathrm{dc}\right) }\left(
\mathrm{v}\right) $ \ [see Eq.~(\ref{e8})]: $\mathcal{N}_{\mathrm{dc}}=
2\gamma _{g}^{\prime}\mathcal{M}_{\mathrm{dc}}$. The limited value $M_{D}^{\left(
\mathrm{dc}\right) }\left( 0\right) $ coincides with the result of the
linear theory~\cite{FisHalPla-1979}.

In the decoupling model~\cite{Vin-1999}, the electron lattice and DL were not
relatively displaced until the instability occurs. It is instructive to investigate
also stability of the WS-DL coupling concerning small relative displacements.
Introducing the displacement parameter $b$ along the $x$-axis, one can find
\[
\frac{F_{D}^{\left( \mathrm{dc}\right) }\left( b\right) }{m_{e}}=-\sum_{%
\mathbf{g}}\Gamma _{g}\frac{g_{x}\omega _{g}^{2}}{g^{2}}\mathcal{M}_{\mathrm{%
dc}}\left( \frac{\mathrm{v}}{\mathrm{v}_{g_{x}}},\frac{\gamma _{g}}{\omega
_{g}}\right) \times
\]%
\begin{equation}
\times \left\{ \left[ 1-\left( \frac{\mathrm{v}}{\mathrm{v}_{g_{x}}}\right)
^{2}\right] \sin \left( g_{x}b\right) +2\frac{\gamma _{g}}{\omega _{g}}\frac{%
\mathrm{v}}{\mathrm{v}_{g_{x}}}\mathrm{\cos }\left( g_{x}b\right) \right\} .
\label{e11}
\end{equation}%
Eqs.~(\ref{e5}) and (\ref{e6}) can be found from Eq.~(\ref{e11}) by setting $b$
to zero. The second term in the  curly brackets is a decreasing function of $b$,
but it is much smaller than the first term except for the limiting case $\mathrm{v}\rightarrow \mathrm{v}_{g_{x}}$.
Considering only a small relative displacement and assuming $\gamma
_{g}/\omega _{g} \ll 1$, one can represent $F_{D}^{\left( \mathrm{dc}\right) }$ as%
\begin{equation}
F_{D}^{\left( \mathrm{dc}\right) }\left( b\right) \simeq -m_{e}\Omega ^{2}b,
\label{e12}
\end{equation}%
where%
\begin{equation}
\Omega ^{2}\left( \mathrm{v}\right) =\sum_{\mathbf{g}}\frac{g_{x}^{2}}{g^{2}}%
\frac{\Gamma _{g}\omega _{g}^{2}\left[ 1-\left( \frac{\mathrm{v}}{\mathrm{v}%
_{g_{x}}}\right) ^{2}\right] }{\left[ 1-\left( \frac{\mathrm{v}}{\mathrm{v}%
_{g_{x}}}\right) ^{2}\right] ^{2}+4\left( \frac{\gamma _{g}}{\omega _{g}}%
\frac{\mathrm{v}}{\mathrm{v}_{g_{x}}}\right) ^{2}}.  \label{e13}
\end{equation}%
In the linear regime $\mathrm{v}\ll \mathrm{v}_{1}^{\left( \mathrm{bc}%
\right) }$ (here $\mathrm{v}_{1}^{\left( \mathrm{bc}\right) }$ is the
smallest value of $\mathrm{v}_{\left\vert g_{x}\right\vert }$), $\Omega
\left( 0\right) $ coincides with the typical frequency of electron
oscillations in the bottom of a surface dimple $\omega _{\mathrm{f}}$ which
is usually much higher than $\omega _{g_{1}}$. Therefore, at $\mathrm{v}<%
\mathrm{v}_{1}^{\left( \mathrm{bc}\right) }$ the quantity $\Omega ^{2}>0$ and
the strong force of Eq.~(\ref{e12}) restores the equilibrium position of the
electron lattice. On the contrary, at $\mathrm{v}>\mathrm{v}_{1}^{\left(
\mathrm{bc}\right) }$ the $\Omega ^{2}\left( \mathrm{v}\right) $ becomes
negative, and the force of Eq.~(\ref{e12}) provides an additional drive
for WS decoupling from the DL.

\subsection{The ac theory of nonlinear effects for a given-current regime}

As noted in the Introduction, the exact solution of Eq.~(\ref{e3}) can be
found only for a given-current regime considering the harmonic dependence of the
WS motion: $u\left( t\right) =u_{a}\sin \left( \omega t\right) $. In this
treatment, the force acting on an electron due to interactions with the DL
can be represented as a sum of harmonics. The first harmonic can be found as~%
\cite{Mon-2018}%
\begin{equation}
F_{D}^{\left( \mathrm{ac}\right) }=-m_{e}\nu _{D}^{\left( \mathrm{ac}\right)
}\mathrm{v}_{a}\cos \left( \omega t\right) +M_{D}^{\left( \mathrm{ac}\right)
}\omega \mathrm{v}_{a}\sin \left( \omega t\right) ,  \label{e14}
\end{equation}%
where the functions $\nu _{D}^{\left( \mathrm{ac}\right) }\left( \mathrm{v}%
_{a},\omega \right) $ and $M_{D}^{\left( \mathrm{ac}\right) }\left( \mathrm{v%
}_{a},\omega \right) $ have the same physical meaning as $\nu _{D}^{\left(
\mathrm{dc}\right) }\left( \mathrm{v}\right) $ and $M_{D}^{\left( \mathrm{dc}%
\right) }\left( \mathrm{v}\right) $ introduced above for the dc conditions
(here $\mathrm{v}_{a}=\omega u_{a}$ is the velocity amplitude). Indeed, in the
right side of Eq.~(\ref{e14}), the first term  represents the nonlinear
kinetic friction (it is proportional to velocity), while the second term
takes into account the dimple inertia (it is proportional to
acceleration).

The expressions for $\nu _{D}^{\left( \mathrm{ac}\right) }$ and $%
M_{D}^{\left( \mathrm{ac}\right) }$ are found as the Fourier transforms of
Eq.~(\ref{e4}):%
\begin{equation}
\frac{M_{D}^{\left( \mathrm{ac}\right) }\left( \mathrm{v}_{a}\right) }{m_{e}}%
=\sum_{\mathbf{g}}\Gamma _{g}\frac{g_{x}^{2}}{g^{2}}\mathcal{M}_{\mathrm{ac}%
}\left( \frac{\mathrm{v}_{a}}{\mathrm{v}_{g_{x}}},\frac{\omega }{\omega _{g}}%
,\frac{\gamma _{g}}{\omega _{g}}\right) ,  \label{e15}
\end{equation}%
\begin{equation}
\nu _{D}^{\left( \mathrm{ac}\right) }\left( \mathrm{v}_{a}\right) =\sum_{%
\mathbf{g}}\Gamma _{g}\frac{g_{x}^{2}}{g^{2}}\omega _{g}\mathcal{N}_{\mathrm{%
ac}}\left( \frac{\mathrm{v}_{a}}{\mathrm{v}_{g_{x}}},\frac{\omega }{\omega
_{g}},\frac{\gamma _{g}}{\omega _{g}}\right) ,  \label{e16}
\end{equation}%
where
\[
\mathcal{M}_{\mathrm{ac}}\left( \mathrm{v}_{a}^{\prime },\omega ^{\prime
},\gamma _{g}^{\prime }\right) =-\frac{2}{\omega ^{\prime }\mathrm{v}%
_{a}^{\prime }}\int\limits_{0}^{\infty }\sin \left( x\right) \mathrm{e}%
^{-\gamma _{g}^{\prime }x}\sin \left( \frac{\omega ^{\prime }x}{2}\right)
\times
\]
\begin{equation}
\times J_{1}\left( 2\frac{\mathrm{v}_{a}^{\prime }}{\omega ^{\prime }}\sin \left(
\frac{\omega ^{\prime }x}{2}\right) \right) dx,\text{ \ }  \label{e17}
\end{equation}%
\[
\mathcal{N}_{\mathrm{ac}}\left( \mathrm{v}_{a}^{\prime },\omega ^{\prime
},\gamma _{g} ^{\prime }\right) =\frac{2}{\mathrm{v}_{a}^{\prime }}%
\int\limits_{0}^{\infty }\sin \left( x\right) \mathrm{e}^{-\gamma
_{g}^{\prime }x}\cos \left( \frac{\omega ^{\prime }x}{2}\right)
\times
\]
\begin{equation}
\times
J_{1}\left( 2\frac{\mathrm{v}_{a}^{\prime }}{\omega ^{\prime }}\sin
\frac{\omega ^{\prime }x}{2}\right) dx\,,  \label{e18}
\end{equation}%
$J_{1}\left( z\right) $ is the Bessel function, $\mathrm{v}_{a}^{\prime }=%
\mathrm{v}_{a}/\mathrm{v}_{g_{x}}$, and $\omega ^{\prime }=\omega /\omega
_{g}$. Other notations are the same as in Eq.~(\ref{e8}). In Eqs.~(\ref{e17})
and (\ref{e18}), one can consider only positive values of the parameter $%
\mathrm{v}_{a}^{\prime }\propto g_{x}$ because $\mathcal{M}_{\mathrm{ac}}$
and $\mathcal{N}_{\mathrm{ac}}$ are independent of the sign of $\mathrm{v}%
_{a}^{\prime }\left( g_{x}\right) $. Alternatively, we can redefine $\mathrm{%
v}_{a}^{\prime }=\mathrm{v}_{a}/\mathrm{v}_{\left\vert g_{x}\right\vert }$.

For the given-current regime ($\mathrm{v}=\mathrm{v}_{a}\cos \omega t$),
the first harmonic of the driving electric field is found from Eq.~(\ref{e3})
as $E\left( t\right) =E_{a}\sin \left( \omega t+\beta \right) $, where
\[
\sin \beta =-\nu _{D}^{\left( \mathrm{ac}\right) }\left[ \left( \nu
_{D}^{\left( \mathrm{ac}\right) }\right) ^{2}+\left( 1+\frac{M_{D}^{\left(
\mathrm{ac}\right) }}{m_{e}}\right) ^{2}\omega ^{2}\right] ^{-1/2}.
\]%
The relation between amplitudes of the field and velocity
\begin{equation}
E_{a}=\frac{m_{e}}{e}\mathrm{v}_{a}\sqrt{\left( \nu _{D}^{\left( \mathrm{ac}%
\right) }\right) ^{2}+\left( 1+\frac{M_{D}^{\left( \mathrm{ac}\right) }}{%
m_{e}}\right) ^{2}\omega ^{2}}  \label{e19}
\end{equation}%
can be considered as the balance equation for the amplitudes, where $\nu
_{D}^{\left( \mathrm{ac}\right) }$ and $M_{D}^{\left( \mathrm{ac}\right) }$
are rather complicated functions of $\mathrm{v}_{a}$, $\omega $ and the
ripplon damping. The right side of Eq.~(\ref{e19}) as a function of $\mathrm{v%
}_{a}$ has a maximum. The balance of forces is impossible if the
field amplitude $E_{a}$ exceeds this maximum. Therefore, this ac treatment
is an extension of the dc decoupling model where the current was fixed to a
constant value.

It is important that in the ac theory, the dimple mass
$M_{D}^{\left( \mathrm{ac}\right) }$ enters the balance of forces equation
for amplitudes, and at high enough frequencies it can affect WS decoupling.

\subsection{Comparison with results of the given-filed regime}

Naturally, there is a question: can we rely on the results of the
given-current treatment when considering a given-field regime? To
analyze this problem let us consider the given-field regime and investigate
the delicate behavior of functions $M_{D}^{\left( \mathrm{ac}\right) }$ and $%
\nu _{D}^{\left( \mathrm{ac}\right) }$ near subharmonic resonances ($\omega \sim \omega _{g}/2$)
established previously~\cite{Mon-2017,Mon-2018} for the inverse treatment.
This frequency range is important for understanding the subharmonic phonon-ripplon
coupling~\cite{Mon-2017}. A comparison of low-frequency ($\omega \rightarrow 0$)
results is presented in the next Subsection.

Of course, in the given-field regime, the equation of motion [Eq.~(\ref{e3})] cannot
be solved exactly. Still, we can expand the exponential
functions entering Eqs.~(\ref{e1}) and (\ref{e2}) in power series restricting
our consideration to the first nonlinear terms.
Now, the driving field is fixed to a harmonic function $E\left(
t\right) =E_{a}\cos \left( \omega t\right) $, while $u\left( t\right) $ can
be represented as a Fourier series%
\begin{equation}
u\left( t\right) =\sum_{k=-\infty }^{\infty }u_{k}\mathrm{e}^{ik\omega t}.
\label{e20}
\end{equation}%
We shall assume that the amplitudes of higher harmonics $\left\vert
u_{k}\right\vert $ decrease fast with $\left\vert k\right\vert $. By
definition, the driving field has only one harmonic ($k=\pm 1$) with $%
E_{1}=E_{-1}=E_{a}/2$.

Using Eq.~(\ref{e20}), we can find the equation of motion of the
coupled WS-DL system for a weak nonlinearity%
\begin{equation}
-\omega ^{2}\mathcal{Z}_{k}u_{k}+\chi _{k}^{\mathrm{NL}}=\frac{eE_{k}}{m_{e}}%
,  \label{e21}
\end{equation}%
where $\chi _{k}^{\mathrm{NL}}$ is the contribution from nonlinear terms%
\[
\chi _{k}^{\mathrm{NL}}=\frac{1}{6}\sum_{\mathbf{g}}\Gamma _{g}\omega
_{g}^{4}g^{2}\frac{g_{x}^{4}}{g^{4}}\left( \frac{1}{\Delta _{k,g}}-\frac{1}{%
\Delta _{0,g}}\right) \sum_{n,l}u_{n}u_{l}u_{k-n-l}+
\]%
\begin{equation}
+\frac{1}{2}\sum_{\mathbf{g}}\Gamma _{g}\omega _{g}^{4}g^{2}\frac{g_{x}^{4}}{%
g^{4}}\sum_{l}u_{k-l}\left( \frac{1}{\Delta _{k-l,g}}-\frac{1}{\Delta _{l,g}}%
\right) \sum_{n}u_{n}u_{l-n},  \label{e22}
\end{equation}%
and we had used the following notations: $\Delta _{k,g}=\omega
_{g}^{2}-k^{2}\omega ^{2}+2ik\omega \gamma _{g}$,
\begin{equation}
\mathcal{Z}_{k}\left( \omega \right) =k^{2}+\sum_{\mathbf{g}}\Gamma _{
g}\left( \frac{\omega _{g}}{\omega }\right) ^{2}\frac{k^{2}\omega
^{2}-2ik\omega \gamma _{g}}{\omega _{g}^{2}-k^{2}\omega ^{2}+2ik\omega
\gamma _{g}}.  \label{e23}
\end{equation}%
The appearance of only cubic terms in $\chi _{k}^{\mathrm{NL}}$ is caused by
the fact that in the integrand of Eq.~(\ref{e4}) we have to expand the sine
function. In the limit $\gamma _{g}\rightarrow 0$, the quantity $\mathcal{Z}%
_{1}\left( \omega \right) $ represents the linear response function $%
\mathcal{Z}_{1}\left( \omega \right) \rightarrow 1+M_{D}^{\left( \mathrm{ac}%
\right) }\left( \omega \right) /m_{e}$ which is frequently used for the
description of electron-ripplon resonances~\cite{MonKon-book,MonSyv-2012} and
phonon-ripplon coupling. Note that in the linear ac theory, the dimple-mass
function $M_{D}^{\left( \mathrm{ac}\right) }\left( \omega \right) $ has a
resonance at $\omega \rightarrow \omega _{g}$, which is the reason for
phonon-ripplon coupling~\cite{FisHalPla-1979}, and it changes its sign for $%
\omega >\omega _{g}$. The negative values of $M_{D}^{\left( \mathrm{ac}%
\right) }\left( \omega \right) $ represent an oscillatory response of the
DL: the WS and DL oscillate out of phase.

Consider the nonlinear equation for the first harmonic $u_{1}$. Assuming
that higher harmonics $u_{k}$ decrease fast with $k$, one can find the first
nonlinear correction to the effective mass function $M_{D}^{\left( \mathrm{ac%
}\right) }\left( \omega \right) $%
\[
\frac{\delta M_{D}}{m_{e}}=\frac{\left\vert u_{1}^{\left( 0\right)
}\right\vert ^{2}}{2}\sum_{\mathbf{g}}\Gamma _{g}\frac{g_{x}^{4}\omega
_{g}^{4}}{g^{2}\omega ^{2}}\times
\]%
\[
\times \mathrm{Re}\left( \frac{\Delta _{-1,g}-\Delta _{2,g}}{\Delta
_{2,g}\Delta _{-1,g}}-3\frac{\omega _{g}^{2}-\Delta _{1,g}}{\Delta
_{1,g}\omega _{g}^{2}}\right) ,
\]%
where $u_{1}^{\left( 0\right) }$ is the solution of the linear equation.
Using the definitions of Eq.~(\ref{e15}) and neglecting ripplon damping, the
dimensionless mass function can be written as%
\begin{equation}
\mathcal{M}_{\mathrm{ac}}\left( \mathrm{v}_{a,0}^{\prime }\right) =\frac{1}{%
1-\left( \omega ^{\prime }\right) ^{2}}\left\{ 1+\frac{3}{2}\frac{\left(
\mathrm{v}_{a,0}^{\prime }\right) ^{2}}{\left[1-\left( 2\omega ^{\prime }\right)
^{2}\right]}\right\}   \label{e24}
\end{equation}%
where $\mathrm{v}_{a,0}^{\prime }=2u_{1}^{\left( 0\right) }\omega /\mathrm{v}%
_{\left\vert g_{x}\right\vert }=u_{a}^{\left( 0\right) }\omega /\mathrm{v}%
_{\left\vert g_{x}\right\vert }$, and $u_{a}^{\left( 0\right) }$ is the
linear approximation for the amplitude of the first harmonic of $u\left(
t\right) $. Already from this equation, one can see that the nonlinear
correction to $\mathcal{M}_{\mathrm{ac}}$ leads to the subharmonic resonance
of the dimple mass at $\omega \rightarrow \omega _{g}/2$, which agrees with
the result found for the given-current regime. Higher nonlinear terms surely
will lead to other subharmonic resonances at $\omega \rightarrow \omega
_{g}/l$ (here $l=3,4...$).

Another nonlinear effect which can be seen from
Eq.~(\ref{e24}) is that the velocity amplitude $\mathrm{v}_{a,0}^{\prime }$
reduces the first harmonic resonance ($\omega \rightarrow \omega _{g}$)
because the second term in curly brackets becomes negative if $\omega
^{\prime }\simeq 1$. Eventually this reduction can change the sign of
resonant variations of $\mathcal{M}_{\mathrm{ac}}\left( \mathrm{v}%
_{a,0}^{\prime },\omega ^{\prime }\right) $.

\begin{figure}[tbp]
\begin{center}
\includegraphics[width=10.5cm]{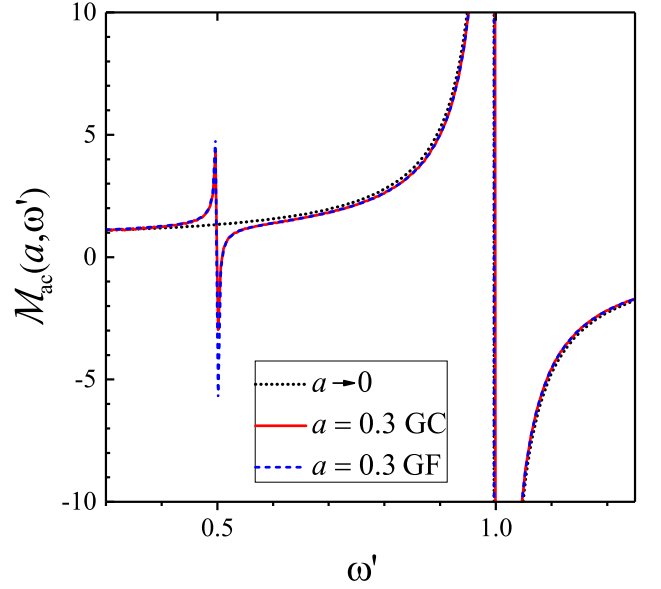}
\end{center}
\caption{The dimensionless dimple-mass function $\mathcal{M}_{\mathrm{ac}}
\left(a, \omega ^{\prime }\right)$ versus frequency $\omega ^{\prime }$
calculated for the given-current (GC) and given-field (GF) regimes at
a fixed value of the nonlinear parameter $a=\left\vert g_{x}\right\vert
u_{a}^{\left( 0\right) }=0.3$.
The exact solution of the GC regime was calculated at $\gamma ^{\prime } _{g}=0.0025$,
while the asymptote of the GF regime [Eq.~(\ref{e24})] has $\gamma ^{\prime } _{g}=0$. }
\label{f1}
\end{figure}

The comparison of the exact solution of the given-current regime [Eq.~(\ref%
{e17})] with the approximation [Eq.~(\ref{e24})] found for the given-field
regime is shown in Fig.~\ref{f1}. Following notations of Ref.~\onlinecite{Mon-2018}%
, we shall use the dimensionless parameter $a=\left\vert g_{x}\right\vert
u_{a}^{\left( 0\right) }\equiv \left\vert g_{x}\right\vert \mathrm{v}%
_{a,0}/\omega $ describing the strength of nonlinearity because it enters
the exponential functions of Eqs.~(\ref{e1}) and (\ref{e2}). The exact result
of Eq.~(\ref{e17}) shown in Fig.~\ref{f1} by the solid (red) line was
calculated for $a=0.3$ and a small but finite damping coefficient ($\gamma
_{g}^{\prime }=0.0025$) because the integrand represents a highly oscillating
function. The dashed (blue) line represents the approximation of Eq.~(\ref%
{e24}) taken at the same value of the nonlinear parameter $a$. One can see
that the approximate form of the given-filed regime is so close to the exact
form of the given-current regime that it is impossible to distinguish the
two lines except for a small vicinity of $\omega ^{\prime }=1/2$ where the
ripplon damping affects the exact solution.

\begin{figure}[tbp]
\begin{center}
\includegraphics[width=10.5cm]{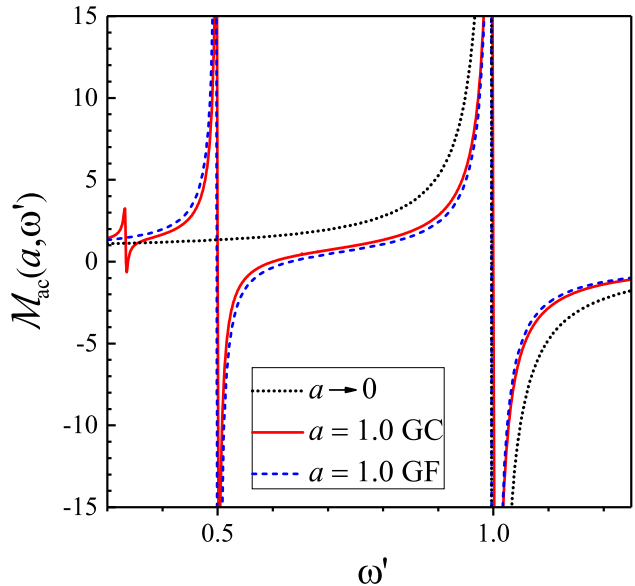}
\end{center}
\caption{The dimensionless dimple-mass function $\mathcal{M}_{\mathrm{ac}}
\left(a, \omega ^{\prime }\right)$ versus frequency $\omega ^{\prime }$
calculated for the given-current (GC) and given-field (GF) regimes at
a fixed value of the nonlinear parameter $a=\left\vert g_{x}\right\vert
u_{a}^{\left( 0\right) }=1$.}
\label{f2}
\end{figure}

At a substantially larger value of the nonlinear parameter $a=1$, the
deviations of the approximate equation of the given-field regime from the
exact result of the given-current regime become noticeable
as indicated in Fig.~\ref{f2}. The approximate form leads to a bit stronger
nonlinear effect. Nevertheless, the approximate form is not far away from
the exact result except for the vicinity of $\omega ^{\prime }=1/3
$, where higher nonlinear terms become important according to the red-solid
line of Fig.~\ref{f2}.

In the given-field regime, finding higher terms of the expansion of
$\mathcal{M}_{\mathrm{ac}}\left( \mathrm{v}_{a,0}^{\prime }\right)$ in power
series is a very difficult task. Still, we can find a better approximation
if the second term in curly brackets of Eq.~(\ref{e24}) is improved as
\[
\frac{3}{2}\frac{\left(
\mathrm{v}_{a,0}^{\prime }\right) ^{2}}{\left[1-\left( 2\omega ^{\prime }\right)
^{2}\right]}
\left[ 1+\frac{5}{4}\frac{\left( \mathrm{v}_{a,0}^{\prime }\right) ^{2}}{%
\left[ 1-\left( 3\omega ^{\prime }\right) ^{2}\right] }\right] .
\]
This rule follows from the expansion of the low-frequency expression for
$\mathcal{M}_{\mathrm{ac}}\left( \mathrm{v}_{a}^{\prime }\right)$ found
for the given-current regime in the next Subsection. This improvement
makes the dashed line very close to the solid line of Fig.~\ref{f2}
even for $a=1$. The algorithm of finding series expansion of Eq.~(\ref{e17})
allows us to calculate the dimple mass at a necessary accuracy even for
zero ripplon damping.

The appearance of maxima of the function $\mathcal{N}_{\mathrm{ac}}$ at
subharmonic frequencies $\omega \rightarrow \omega _{g}/l$ (here $l=2,3,...$) is quite obvious
even from Eq.~(\ref{e22}) obtained for the given-field regime. Therefore, we
conclude that the results obtained for $M_{D}^{\left( \mathrm{ac}\right)
}\left( \mathrm{v}_{a}\right) $ and $\nu _{D}^{\left( \mathrm{ac}\right)
}\left( \mathrm{v}_{a}\right) $ in the given-current regime can quite well
describe the respective quantities measured in the given-field regime at
medium frequencies $\omega < \omega _{g_{1}}$. The case of a much low frequency
$\omega \ll \omega _{g_{1}}$ is considered below.

\section{Low-frequency limit of the nonlinear ac transport}

Experiments on nonlinear WS transport over liquid helium usually were
conducted for an ac driving potential. In this case, field-velocity
characteristics were presented for time-averaged quantities or amplitudes.
Therefore, we expect that results obtained using a nonlinear dc model may
not be applied to the data found in an ac experiment.
Partly this problem was analyzed in Refs.~\cite{MonKon-2009,Mon-2018}. Here
we would like to investigate the low-frequency limit of the nonlinear ac
transport of the WS over superfluid $^{4}\mathrm{He}$ at $T<0.4\,\mathrm{K}$. In this case, ripplon
damping is extremely small and the ratio $\gamma _{g_{1}}/\omega _{g_{1}}<10^{-4}$
(here $g_{1}$ is the smallest reciprocal lattice vector).
At the same time the frequency of the driving potential in experiments on
the WS transport is rather low, and a typical value of
the ratio $\omega /\omega _{g_{1}}\sim 10^{-3}$. Therefore, it is important
to find the asymptotic behavior of field-velocity characteristics of the ac WS
transport in the limit of low frequencies and damping.

Firstly, consider the limiting case of zero damping. Then, the low-frequency
asymptote of the dimensionless function $\mathcal{M}_{\mathrm{ac}%
}\left( \mathrm{v}_{a}^{\prime },\omega ^{\prime },\gamma _{g}^{\prime
}\right) $ defined above in Eq.~(\ref{e17}) can be found in an analytical form%
\begin{equation}
\mathcal{M}_{\mathrm{ac}}\left( \mathrm{v}_{a}^{\prime }\right) \rightarrow
\frac{\theta \left( 1-\left\vert \mathrm{v}_{a}^{\prime }\right\vert \right)
}{\left[ 1-\left( \mathrm{v}_{a}^{\prime }\right) ^{2}\right] ^{3/2}},
\label{e25}
\end{equation}%
where we use the absolute-value symbol because the definition of $\mathrm{v}%
_{a}^{\prime }$ contains $g_{x}$ which can change its sign. One can notice
that this equation differs substantially from the result of the dc theory
given in Eq.~(\ref{e10}) and taken at $\gamma _{g}^{\prime }=0$. The
difference is caused by the fact that $\mathcal{M}_{\mathrm{ac}}\left(
\mathrm{v}_{a}^{\prime }\right) $ is a time-averaged quantity (a Fourier transform).

It is important to note that expanding
Eq.~(\ref{e25}) in $\left( \mathrm{v}_{a}^{\prime }\right) ^{2}$ and
neglecting the terms of a higher power we obtain the result which coincides
with Eq.~(\ref{e24}) found for the given-field regime and taken at
$\omega ^{\prime }=0$. Thus, we proved the agreement (even numerical) between
the solutions found in two opposite regimes in the wide range of frequencies
which includes the limit $\omega \rightarrow 0$. The result of Eq.~(\ref{e25})
indicates also that a dc transport model cannot always be applied to the
time-averaged quantities or amplitude data of an ac transport experiment.
Another important consequence of Eq.~(\ref{e25}) is that it allows us
to formulate the algorithm for the series expansion of Eq.~(\ref{e17}) in
$\left( \mathrm{v}_{a}^{\prime }\right) ^{2}$
valid for a finite frequency and zero ripplon damping, as discussed in the
preceding Subsection.

\begin{figure}[tbp]
\begin{center}
\includegraphics[width=10.5cm]{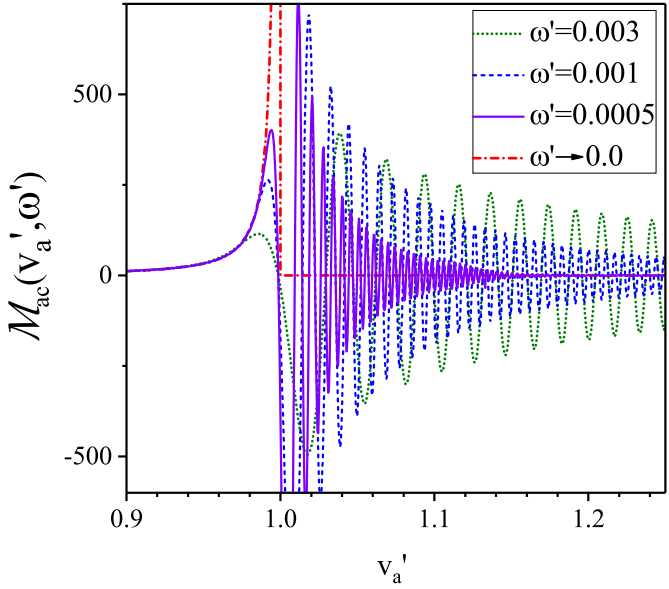}
\end{center}
\caption{The dimensionless dimple-mass function $\mathcal{M}_{\mathrm{ac}}
\left(\mathrm{v}_{\mathrm{a}} ^{\prime }, \omega ^{\prime }\right)$ versus the velocity
amplitude $\mathrm{v}_{\mathrm{a}} ^{\prime }$
calculated for different frequencies. The damping parameter $\gamma _{g}^{\prime }$
is fixed to 0.0025.}
\label{f3}
\end{figure}

It is instructive to see how $\mathcal{M}_{\mathrm{ac}}\left( \mathrm{v}%
_{a}^{\prime },\omega ^{\prime },\gamma _{g}^{\prime }\right) $ of Eq.~(\ref%
{e17}) approaches the asymptote given in Eq.~(\ref{e25}) with lowering
frequency $\omega $. The results of calculations based on Eq.~(\ref{e17}) are
shown in Fig.~\ref{f3} together with the asymptote. We found that at low
velocities $\mathrm{v}_{a}^{\prime }<1$ and $\omega ^{\prime }$ the exact
form of $\mathcal{M}_{\mathrm{ac}}\left( \mathrm{v}_{a}^{\prime },\omega
^{\prime },\gamma _{g}^{\prime }\right) $ coincides with the asymptote
indicated by the red dash-dotted line. Deviations from this line, which
appear in the vicinity of the BC condition $\mathrm{v}_{a}\simeq \mathrm{v}%
_{\left\vert g_{x}\right\vert }$, increase strongly with $\omega ^{\prime
} $. One can see that the BC threshold for the dimple mass is strongly
affected by the finite frequency of the driving field. At a fixed $\gamma
_{g}^{\prime }=0.0025$ the maximum of the threshold decreases with
increasing $\omega ^{\prime }$. A remarkable behavior of the mass function $%
\mathcal{M}_{\mathrm{ac}}\left( \mathrm{v}_{a}^{\prime }\right) $ is seen
above the BC threshold: in this range, it approaches the asymptote of
Eq.~(\ref{e25}) in a highly oscillating way. Moreover, at maxima of these
oscillations, the dimple mass can be even larger than at the BC maximum. This
means that above the BC threshold the WS and DL go through multiple
decoupling and recoupling processes.

\begin{figure}[tbp]
\begin{center}
\includegraphics[width=10.5cm]{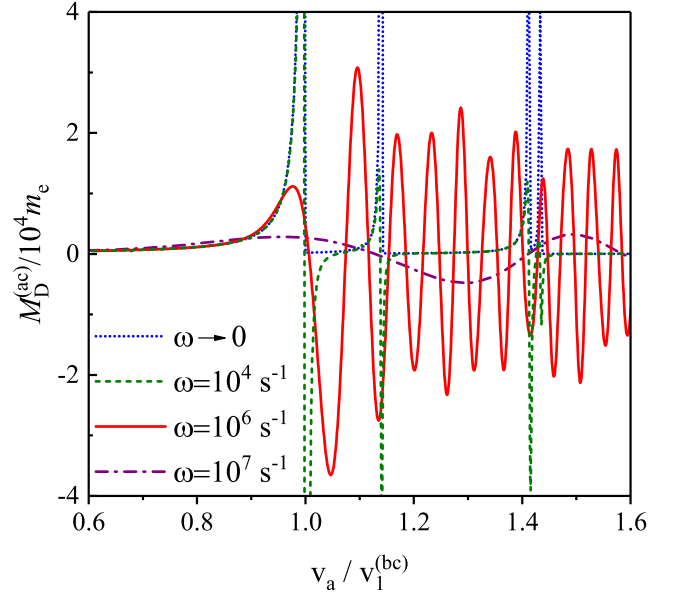}
\end{center}
\caption{The dimple-mass $M_{D}^{\left( \mathrm{ac}\right) }$ normalized
versus the velocity amplitude $\mathrm{v}_{\mathrm{a}}$
calculated for $n_{e}=5\cdot 10^{8}\,\mathrm{cm}^{-2}$, $T=0.15\,\mathrm{K}$
and different frequencies. The damping parameter
$\gamma _{g_{1}}/\omega _{g_{1}}$ is fixed to $0.003$.}
\label{f4}
\end{figure}

The BC threshold exists not only for the smallest $g$; it exists also for
all other reciprocal lattice vectors which have a nonzero component along
the velocity vector. Therefore, in the limiting case $\omega ^{\prime
}\rightarrow 0$, the dimple mass $M_{D}^{\left( \mathrm{ac}\right) }$ has many
sharp peaks as illustrated in Fig.~\ref{f4} by the blue-dotted line.
At a rather low frequency $\omega =10^{4}\,\mathrm{s}^{-1}$ and $\gamma
_{g_{1}}^{\prime }=0.003$, the calculated line (olive-dashed) has finite
maxima and negative minima just above the BC conditions. It should be noted
that negative values of $M_{D}^{\left( \mathrm{ac}\right) }$ simply mean the
oscillatory response of the DL similar to that discussed above just after
Eq.~(\ref{e23}). For the substantially higher frequency $\omega =10^{6}%
\,\mathrm{s}^{-1}$, the position of maxima (except for the first one) and
decoupling processes have no relation to the BC conditions. The further
increase in frequency ($\omega =10^{7}\,\mathrm{s}^{-1}$) strongly suppresses
the BC effect resulting in a smooth line (dash-dotted purple).

\begin{figure}[tbp]
\begin{center}
\includegraphics[width=10.5cm]{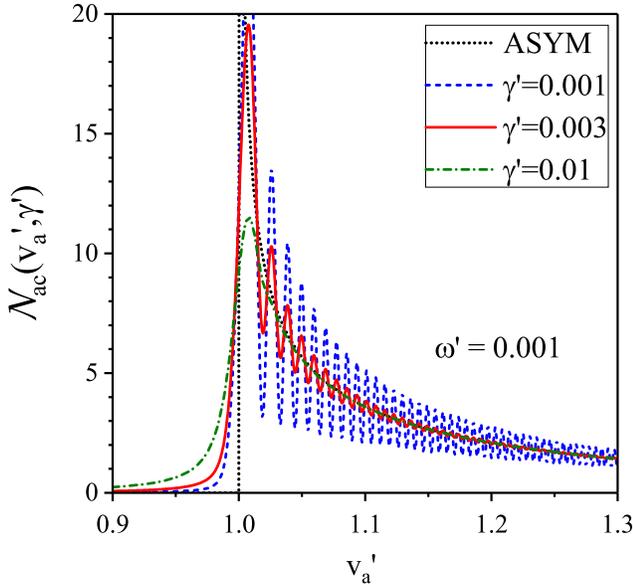}
\end{center}
\caption{The dimensionless function $\mathcal{N}_{\mathrm{ac}}
\left(\mathrm{v}_{\mathrm{a}} ^{\prime }, \gamma ^{\prime }\right)$ versus the velocity
amplitude $\mathrm{v}_{\mathrm{a}} ^{\prime }$
calculated for a fixed frequency $\omega ^{\prime}=0.001$ and different values of the
damping parameter $\gamma ^{\prime}=\gamma _{g}/\omega _{g}$. The black dotted line
represents the asymptote (ASYM) of Eq.~(\ref{e26}).
}
\label{f5}
\end{figure}

\begin{figure}[tbp]
\begin{center}
\includegraphics[width=10.5cm]{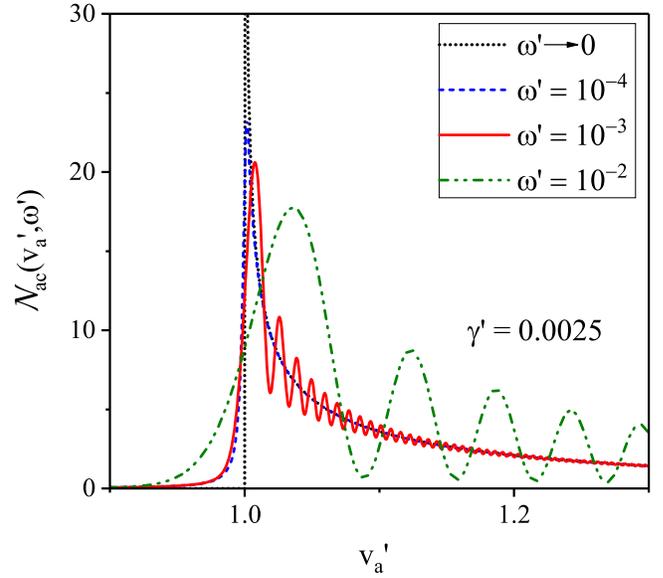}
\end{center}
\caption{The dimensionless function $\mathcal{N}_{\mathrm{ac}}
\left(\mathrm{v}_{\mathrm{a}} ^{\prime }, \omega ^{\prime }\right)$ versus the velocity
amplitude $\mathrm{v}_{\mathrm{a}} ^{\prime }$
calculated for a fixed damping parameter and different frequencies $\omega ^{\prime}$.
Other notations are the same as in Fig.~\ref{f5}.
}
\label{f6}
\end{figure}

Consider now the nonlinear kinetic friction described by the functions $\nu
_{D}^{\left( \mathrm{ac}\right) }$ and $\mathcal{N}_{\mathrm{ac}}$. In the
limiting cases $\omega ^{\prime }\rightarrow 0$ and $\gamma _{g}^{\prime }=0$%
, the asymptotic behavior of $\mathcal{N}_{\mathrm{ac}}\left( \mathrm{v}%
_{a}^{\prime }\right) $ defined in Eq.~(\ref{e18}) is found as
\begin{equation}
\mathcal{N}_{\mathrm{ac}}\left( \mathrm{v}_{a}^{\prime }\right) \simeq \frac{%
2\theta \left( \left\vert \mathrm{v}_{a}^{\prime }\right\vert -1\right) }{%
\left( \mathrm{v}_{a}^{\prime }\right) ^{2}\sqrt{\left( \mathrm{v}%
_{a}^{\prime }\right) ^{2}-1}}.  \label{e26}
\end{equation}%
According to this equation, the BC peak of $\mathcal{N}_{\mathrm{ac}}\left(
\mathrm{v}_{a}^{\prime }\right) $ has a strongly asymmetric shape which is
in contrast with the result of the dc model given in Eq.~(\ref{e8}). A comparison
with the given-field regime cannot be given because $\mathcal{N}_{\mathrm{ac}}\left(
\mathrm{v}_{a}^{\prime }\right) $ of Eq.~(\ref{e26}) differs from zero only at
high-velocity amplitudes $\left\vert \mathrm{v}_{a}^{\prime }\right\vert \geq 1$. The
evolution of the exact function $\mathcal{N}_{\mathrm{ac}}\left( \mathrm{v}%
_{a}^{\prime }\right) $ calculated for a fixed $\omega ^{\prime }=0.001$ and
several values of the damping parameter $\gamma _{g}^{\prime }$ is shown in
Fig.~\ref{f5}. An increase in $\gamma _{g}^{\prime }$ reduces the
BC-threshold maximum and the amplitude of oscillations in the tail region. A
different behavior of $\mathcal{N}_{\mathrm{ac}}\left( \mathrm{v}%
_{a}^{\prime }\right) $ is obtained for a fixed $\gamma _{g}^{\prime }=0.0025
$ with increasing the frequency parameter $\omega ^{\prime }$ as shown in
Fig.~\ref{f6}. The BC maximum is reduced by an increase in $\omega ^{\prime }
$, but the amplitude of oscillations and their period in the tail region
become larger.

\begin{figure}[tbp]
\begin{center}
\includegraphics[width=10.5cm]{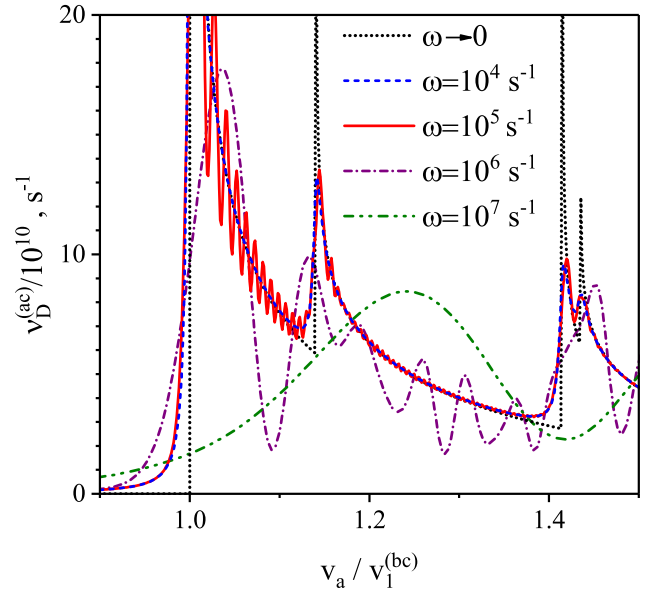}
\end{center}
\caption{The effective collision frequency
caused by electron interaction with a surface dimple $\nu_{D}^{\left( \mathrm{ac}\right) }$
versus the velocity amplitude $\mathrm{v}_{\mathrm{a}}$
calculated for $n_{e}=5\cdot 10^{8}\,\mathrm{cm}^{-2}$, $T=0.15\,\mathrm{K}$,
and different frequencies. The damping parameter
$\gamma _{g_{1}}/\omega _{g_{1}}$ is fixed to $0.003$.}
\label{f7}
\end{figure}

The typical behavior of $\nu _{D}^{\left( \mathrm{ac}\right) }\left( \mathrm{%
v}_{a}\right) $ is illustrated in Fig.~\ref{f7} for different frequencies $%
\omega \ll \omega _{g_{1}}$. It is remarkable that above the BC threshold $%
\nu _{D}^{\left( \mathrm{ac}\right) }\left( \mathrm{v}_{a}\right) $ is
finite (except for BC points of larger $\mathbf{g}$) even for zero damping
(black dotted line). This greatly distinguishes the ac theory from the dc
model, where $\nu _{D}^{\left( \mathrm{dc}\right) }\left( \mathrm{v}%
_{a}\right) $ is zero except for the BC $\delta $-peaks if $\gamma _{g}=0$.
The increase of frequency up to $10^{4}\,\mathrm{s}^{-1}$ (blue dashed line)
makes the transition through the BC conditions smooth. Away from the BC critical points,
this line follows the asymptote of Eq.~(\ref{e26}). At even higher $\omega
=10^{5}\,\mathrm{s}^{-1}$ the line (red solid) oscillates near the the
asymptotic line of the limiting case $\omega =0$. The amplitude of these
oscillations is substantial only in the vicinity of the first BC peak, and
it decreases for higher velocities where this solid line is close to the
blue-dashed line and the asymptote of Eq.~(\ref{e26}). The higher
frequency $\omega =10^{6}\,\mathrm{s}^{-1}$ (purple dash-dotted line) makes
the oscillations more pronounced and shifts strongly the BC maximum to the
right. At $\omega =10^{7}\,\mathrm{s}^{-1}$, (olive dash-dot-dotted line) the
BC threshold is practically destroyed.

\section{Discussions and Conclusions}

The results obtained for the low-frequency limit indicate that the BC
emission of surface waves by the WS remains to be an
important feature of electron transport on superfluid helium even under ac
conditions. For low frequencies ($10^{4}\,\mathrm{s}^{-1}\leq \omega \leq 10^{5}%
\,\mathrm{s}^{-1}$), the dimple mass $M_{D}^{\left( \mathrm{ac}\right) }$ and
the effective collision frequency $\nu _{D}^{\left( \mathrm{ac}\right) }$
caused by electron interaction with a surface dimple increase sharply when
velocity amplitude $\mathrm{v}_{a}$ approaches $\mathrm{v}_{1}^{\left(
\mathrm{bc}\right) }$. Therefore, the BC effect should restrict the velocity
and current amplitudes with a moderate increase of the amplitude of the
driving electric field. The typical values of $\nu _{D}^{\left( \mathrm{ac}%
\right) }$ induced by the motion of surface dimples and shown in Fig.~\ref{f7} are
much larger than the rate of electron collisions with thermally excited
ripplons $\nu _{e}$. Therefore, the ordinary drag force $-m_{e}\nu _{e}%
\mathrm{v}_{a}\cos \left( \omega t\right) $ acting on an electron can be
neglected in the force balance equation.

An increase in the frequency of the signal differently affects the BC
maxima of $M_{D}^{\left( \mathrm{ac}\right) }\left( \mathrm{v}_{a}\right) $
and $\nu _{D}^{\left( \mathrm{ac}\right) }\left( \mathrm{v}_{a}\right) $: in
the first case, they are displaced left to lower velocity amplitudes, while
for $\nu _{D}^{\left( \mathrm{ac}\right) }\left( \mathrm{v}_{a}\right) $
they are displaced right to higher $\mathrm{v}_{a}$. Another important
distinction to be noticed here is that the BC peaks of $M_{D}^{\left(
\mathrm{ac}\right) }$ and $\nu _{D}^{\left( \mathrm{ac}\right) }$ have
opposite asymmetries (with respect to $\mathrm{v}_{a}=\mathrm{v}_{1}^{\left(
\mathrm{bc}\right) }$) which become more prominent at low driving
frequencies. For the first BC resonance, with $\omega \rightarrow 0$ and $%
\gamma _{g}\rightarrow 0$, these quantities approach zero at the opposite sides with
respect to the critical point $\mathrm{v}_{a}=\mathrm{v}_{1}^{\left(
\mathrm{bc}\right) }$. Thus, the behaviors of the major transport properties
of the WS obtained here for ac conditions are in contrast with the simple treatment
of a stationary motion ($\mathrm{v}=\mathrm{const}$) of electrons~\cite{Vin-1999}.

Another interesting effect obtained above is the very high values of the
dimple mass $M_{D}^{\left( \mathrm{ac}\right) }\left( \mathrm{v}_{a}\right) $
appeared at the BC maximum and above it ($\mathrm{v}_{a}>\mathrm{v}_{1}^{\left(
\mathrm{bc}\right) }$) at oscillation maxima. These values are about two
orders of magnitude (!) larger than the equilibrium dimple mass $%
M_{D}^{\left( \mathrm{ac}\right) }\left( 0\right) $. Therefore, the dimple
mass should be taken into account in the balance of forces equation even for
low frequencies $\omega \ll \omega _{g_{1}}$.

\begin{figure}[tbp]
\begin{center}
\includegraphics[width=10.0cm]{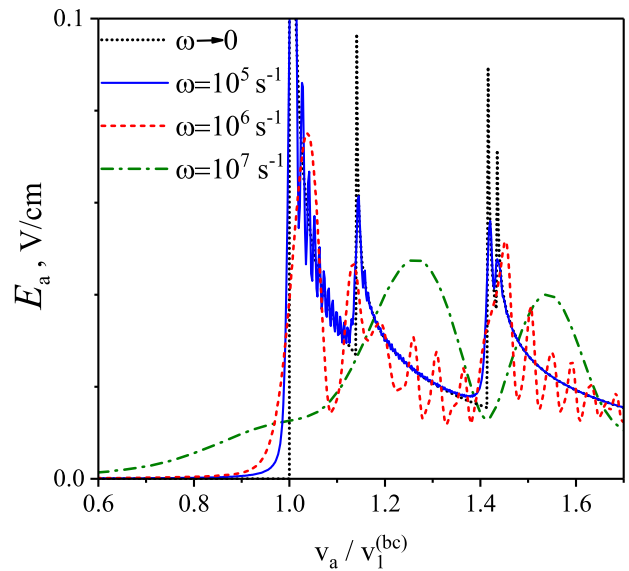}
\end{center}
\caption{Field-velocity characteristics for amplitudes calculated under ac conditions
with different frequencies. Other parameters are the same as in Fig.~\ref{f7}.
 }
\label{f8}
\end{figure}

To obtain the field-velocity characteristic for the ac driving we have used
the balance-of-amplitudes equation given in Eq.~(\ref{e26}). The results of
the calculations are shown in Fig.~\ref{f8}. Since we are considering the
low-frequency limit, the dependence $E_{a}\left( \mathrm{v}_{a}\right) $ is
very similar to the dependence $\nu _{D}^{\left( \mathrm{ac}\right) }\left(
\mathrm{v}_{a}\right) $ illustrated in Fig.~\ref{f7}. Important distinctions
are found in the region $\mathrm{v}_{a}<\mathrm{v}_{1}^{\left( \mathrm{bc}%
\right) }$. One can see that in
this region an increase in the frequency of the signal strongly affects the
field amplitude $E_{a}$ related to a velocity amplitude $\mathrm{v}%
_{a}$. In the opposite region, $E_{a}\left( \mathrm{v}_{a}\right) $ varies
in a quite oscillating way. Still, for the typical value $\omega =10^{5}%
\,\mathrm{s}^{-1}$, the amplitude of oscillations is small, and at average $%
E_{a}\left( \mathrm{v}_{a}\right) $ is remarkably close to the line calculated for the
limiting case $\omega \rightarrow 0$ under ac conditions.

The results presented in Fig.~\ref{f8} are found for amplitudes of the major harmonics
of the electric field and velocity. For a strong nonlinearity, the actual
dependence of $\mathrm{v}\left( t\right)$ can differ substantially from the simple
cosine function because of the presence of higher harmonics appeared in the given-field regime.
These harmonics can affect the shape of current oscillations, as was found in the
time-resolved measurements~\cite{ZouKonRee-2021}. In this experiment, at a strong driving potential amplitude,
the current oscillations obtain a shape with long-flat extrema because of the BC effect (the
velocity amplitude cannot exceed $\mathrm{v}_{1}^{\left( \mathrm{bc}\right) }$).
Then, at substantially larger driving amplitudes the cosine-like shape of the current is almost
restored with some remarkable oscillatory features observed in the vicinity of
current extrema~\cite{ZouKonRee-2021}.

For driving frequencies $\omega \ll \omega _{g_{1}}$, the DL adiabatically follows the WS motion changing its
shape with increasing velocity. In this regime, WS-DL instability can happen according to
the decoupling model. Regarding possible sliding of the WS over the nearly unchanged DL, the analysis of Eqs.~(\ref{e1}) and Eqs.~(\ref{e2}) indicates that it can happen if electron acceleration is high enough.
The driving field should be changing fast within a time scale smaller than
$\omega _{g_{1}}^{-1}$, which makes the process highly nonadiabatic. It should be noted
that in experiments on WS decoupling~\cite{ReeSheKon-2020}, the time scale of the driving field increase
was comparable with the inverse of the ripplon frequency. At such conditions, the sliding model
can be relevant to the sharp velocity jump observed.

In summary, we have investigated the ac nonlinear transport of the 2D Wigner solid
over the free surface of superfluid $^{4}\mathrm{He}$ employing two different approaches. The
first approach is formulated for the given-field regime. In this case, the nonlinear
integro-differential equation of motion is solved approximately taking into account
only the nonlinear terms of the lowest order. In the second approach, the current
is assumed to be fixed to a simple harmonic function (the given-current regime), while
the major harmonic of the driving field is found in an exact form and expressed in terms
of the nonlinear dimple mass function and the effective collision frequency caused by
electron interaction with the DL. Remarkably, for the lowest nonlinear
approximation, both approaches lead to the same results in a wide range of driving
frequencies. This means that the exact results found employing the given-current regime
can be used for the description of the time-averaged measurements performed for the given-field
regime as well.

Under ac driving conditions, we found that the dimple mass and the effective collision
frequency as functions of the velocity amplitude are increased strongly by the BC effect
in a way which differs from that established previously using dc models. The BC maxima
of the field-velocity characteristic calculated for amplitudes obtain an asymmetrical shape,
whose parameters are strongly affected by the frequency of the signal and ripplon damping.
For typical driving frequencies (lower or about $10^{5}\,\mathrm{s}^{-1}$), the field-velocity
characteristic remarkably varies near a universal line calculated for the limiting case
$\omega \rightarrow 0$ and zero damping. The nonlinear dimple mass has only a weak influence
on the field-velocity characteristics. Still, in the given-current regime, it can be very
large even far above the BC threshold.

\end{document}